\documentclass[twocolumn,showpacs]{revtex4}

\usepackage{graphicx}
\usepackage{dcolumn}
\usepackage{bm}

\begin{document}

\title{\Large Wormhole and its Analogue in Brane World}

\author{\bf~Subenoy Chakraborty\footnote{schakraborty@math.jdvu.ac.in}
and~Tanwi~Bandyopadhyay}

\affiliation{Department of Mathematics, Jadavpur University,
Kolkata-32, India.}

\date{\today}

\begin{abstract}
In Einstein gravity, for an inhomogeneous phantom energy
distribution having linear equation of state (but anisotropic),
there exists simple exact solution for spherically symmetric space
time describing a wormhole. At infinity, the space time is not
asymptotically flat and possesses a regular cosmological Killing
horizon with an infinite area. In this work, we have shown that,
this wormhole solution is also possible in brane world for various
matter distribution, which are not necessarily phantom in nature.
\end{abstract}


\maketitle

Wormholes are usually defined as smooth bridges between different
universes or topological handles (i.e, throats, having no horizons)
between remote parts of a single universe. Earlier, wormholes are
purely of theoretical curiosity [1,~2], but recently this
theoretical aspect has gained significant importance due to the
present accelerating phase of our universe [3,~4]. There is a nice
similarity between this theoretical phenomenon and the recent
observational aspects. A traversable wormhole is supported by so
called exotic matter with a negative pressure ($p<0$) and violation
of the null energy condition (i.e, $\rho+p<0$) at least in a
neighbourhood of the wormhole throat [5]. On the other hand, the
observed acceleration of the universe is due to a hypothetical dark
energy, violating the strong energy condition (i.e, $\rho+3p<0$).
A useful choice of dark energy is the phantom energy [6] having
equation of state, $p=-\omega\rho$, $\omega>1$ (thereby violates
null energy condition). However, there is one basic difference
between the above two issues--- in static wormhole, the matter
distribution depends on the spatial coordinates while in cosmology,
the matter density, pressure, are usually functions of temporal
coordinates.\\

In recent years wormholes have been discussed in brane world
scenario, an interesting concept in which, our universe is a 3-brane
embedded in a five dimensional bulk [7,~8,~9]. Bronnikov and Kim
[10] have obtained a class of static and spherically symmetric
solutions in vacuum brane and bulk Weyl effects support the
wormhole. Very recently, Lobo [11] has given a general formulation
for brane world wormholes and has presented two possible wormhole
configurations with dust and perfect fluid having linear equation of
state, as the brane matter.\\

Here we shall show that, wormhole supported by phantom energy, as
described by Zaslavskii [12], can be considered as a possible
wormhole solution in brane scenario for various matter distribution
in the brane and they are not necessarily violating the null energy
conditions.\\

In Schwarzschild coordinates, the spherically symmetric metric
describes that, traversable wormhole can be cast in the form
[13,~14]

\begin{equation}
ds^{2}=-e^{2\Phi}dt^{2}+\frac{dr^{2}}{1-\frac{b(r)}{r}}+r^{2}d{\Omega_{2}}^{2}
\end{equation}

where $\Phi(r)$ is called the redshift function and $b(r)$ stands
for the shape function. In an orthonormal reference frame, the non
vanishing components of the Einstein field equations are

\begin{equation}
\rho(r)=\frac{1}{8\pi G}\frac{b'}{r^{2}}
\end{equation}

\begin{equation}
p_{r}(r)=\frac{1}{8\pi
G}\left[\frac{2}{r}\left(1-\frac{b(r)}{r}\right)\Phi'
-\frac{b}{r^{3}}\right]
\end{equation}

\begin{eqnarray*}
p_{t}(r)=\frac{1}{8\pi
G}\left(1-\frac{b(r)}{r}\right)\left[\Phi''+\Phi'\left(\Phi'
+\frac{1}{r}\right)\right]
\end{eqnarray*}
\begin{equation}
-\frac{(b'r-b)}{2r^{2}}\left(\Phi' +\frac{1}{r}\right)
\end{equation}

in which $\rho(r)$ is the energy density, $p_{r}(r)$ is the radial
pressure and $p_{t}(r)$ is the lateral pressure. The conservation of
the stress-energy tensor gives

\begin{equation}
p_{r}'=\frac{2}{r}(p_{t}-p_{r})-(\rho+p_{r})\Phi'
\end{equation}

Let us suppose that our matter source is described by the phantom
energy with equation of state [14]

\begin{equation}
p_{r}=-\omega_{1}\rho~~~~\text{and}~~~~p_{t}=\omega_{2}\rho,~~~~\omega_{1}>1
\end{equation}

Then Zaslavskii [12] has obtained a simple solution with the choice
$\omega_{2}=\frac{\omega_{1}-1}{4}>0$ as

\begin{eqnarray*}
b=r_{0}+d(r-r_{0}),~~e^{2\Phi}=\frac{r_{1}}{r},~~\rho=\frac{d}{8\pi
r^{2}},
\end{eqnarray*}
\begin{equation}
p_{r}=-\frac{1}{8\pi r^{2}},~~p_{t}=\frac{1-d}{32\pi r^{2}}
\end{equation}

where $r_{0}$ (throat radius) and $r_{1}$ are arbitrary constants
and $d=\frac{1}{\omega_{1}}<1$. Note that as $b'(r_{0})=d<1$, so
flaring of the throat is satisfied. Further, if we demand as usual
$\omega_{2}\leq1$, then $\omega_{1}$ is restricted by the
inequality\\

~~~~~~~~~~~~~~~~~~~~~~$1<\omega_{1}\leq5$\\

The proper radial distance is related to the shape function by

\begin{eqnarray*}
l(r)=\pm{\int_{r_{0}}}^{r}\frac{dr'}{\sqrt{1-\frac{b(r')}{r'}}}
=\pm\frac{1}{\sqrt{1-d}}\left\{\sqrt{r(r-r_{0})}\right.
\end{eqnarray*}
\begin{equation}
\left.+ln\left[\sqrt{\frac{r}{r_{0}}}+\sqrt{\frac{r}{r_{0}}-1}\right]\right\}
\end{equation}

where '$\pm$' signs stand for upper and lower part of the wormhole
or universe. The metric (1) is not asymptotically flat for the
solution (7), rather it can be glued to the external Schwarzschild
solution at some finite radial distance.\\

In Randall-Sundrum type-II brane model [8], the generalized Einstein
equation on the brane has the form [9,~15 and for review see 16]

\begin{equation}
G_{\mu\nu}=\kappa^{2}T_{\mu\nu}+\frac{6\kappa^{2}}{\lambda}~\Pi_{\mu\nu}
-\Lambda~g_{\mu\nu}-\varepsilon_{\mu\nu}
\end{equation}

with $\Lambda=\frac{1}{2}(\Lambda_{5}+\kappa^{2}\lambda)$.\\

Here $\Lambda$ and $\Lambda_{5}$ are respectively the cosmological
constants on the brane and bulk and $\lambda$ is the
brane tension (vacuum energy). The matter confined to the brane is
described by the energy-momentum tensor $T_{\mu\nu}$ (such that
$T_{AB}~n^{B}=0$) and the correction term $\Pi_{\mu\nu}$ (the local
effects of the bulk arising from the brane extrinsic curvature)
which is quadratic in $T_{\mu\nu}$, has the expression

\begin{equation}
\Pi_{\mu\nu}=\frac{1}{12}TT_{\mu\nu}-\frac{1}{4}T_{\mu\alpha}T_{\nu}^{\alpha}
+\frac{1}{8}g_{\mu\nu}\left(T_{\alpha\beta}T^{\alpha\beta}-\frac{1}{3}T^{2}\right)
\end{equation}

with $T=T_{\mu}^{\mu}$.\\

The other correction term $\varepsilon_{\mu\nu}$ denotes the
nonlocal bulk effects and is the projection of the five dimensional
Weyl tensor on the brane, i.e,

\begin{equation}
\varepsilon_{\mu\nu}=~^{(5)}C_{ABCD}~n^{B}n^{D}\delta_{\mu}^{A}\delta_{\nu}^{C}
\end{equation}

with $\varepsilon_{\mu}^{\mu}=0$ (traceless).\\

Due to static and spherically symmetric nature of the wormhole
metric (1), we consider an isotropic fluid on the brane, i.e,

\begin{equation}
T_{\mu\nu}=diag(\rho,p,p,p)
\end{equation}

where $\rho(r)$ is the energy density and $p(r)$ is the isotropic
pressure. As a result, the nonlocal bulk effects contribute in the
form of an effective anisotropic fluid as

\begin{equation}
\varepsilon_{\mu\nu}=diag~[\epsilon(r),\sigma_{r}(r),\sigma_{t}(r),\sigma_{t}(r)]
\end{equation}

Thus the generalized Einstein equation (9) on the brane for the metric (1)
with matter distribution given by (12) and (13), is the modified version of
the equations (2)-(4), where the left hand sides viz. $\rho$, $p_{r}$ and $p_{t}$
are replaced by $\rho^{eff}(r)$, $p_{r}^{eff}(r)$ and $p_{t}^{eff}(r)$ respectively
i.e, (choosing $8\pi G=1$)

\begin{equation}
\frac{b'}{r^{2}}=\rho^{eff}(r)=\rho\left(1+\frac{\rho}{2\lambda}\right)
-\frac{\epsilon}{8\pi}
\end{equation}

\begin{eqnarray*}
\frac{2}{r}\left(1-\frac{b}{r}\right)~\Phi'-\frac{b}{r^{3}}=p_{r}^{eff}(r)
\end{eqnarray*}
\begin{equation}
=p\left(1+\frac{\rho}{\lambda}\right)+\frac{\rho^{2}}{2\lambda}-\frac{\sigma_{r}}{8\pi}
\end{equation}

and

\begin{eqnarray*}
\left(1-\frac{b}{r}\right)\left[\Phi''+\Phi'\left(\Phi'+\frac{1}{r}\right)\right]
-\frac{(b'r-b)}{2r^{2}}
\left(\Phi'+\frac{1}{r}\right)
\end{eqnarray*}
\begin{equation}
=p_{t}^{eff}(r)=p\left(1+\frac{\rho}{\lambda}\right)+\frac{\rho^{2}}{2\lambda}
-\frac{\sigma_{t}}{8\pi}
\end{equation}

Now if we claim that the solution (7) is also the solution of
equations (14)-(16) (with matter components replaced by their
effective components), then we must have

\begin{equation}
\rho\left(1+\frac{\rho}{2\lambda}\right)-\frac{\epsilon}{8\pi}=\frac{d}{8\pi r^{2}}
\end{equation}

\begin{equation}
p\left(1+\frac{\rho}{\lambda}\right)+\frac{\rho^{2}}{2\lambda}-\frac{\sigma_{r}}{8\pi}
=-\frac{1}{8\pi r^{2}}
\end{equation}

\begin{equation}
p\left(1+\frac{\rho}{\lambda}\right)+\frac{\rho^{2}}{2\lambda}-\frac{\sigma_{t}}{8\pi}
=\frac{(1-d)}{32\pi r^{2}}
\end{equation}

Also the tracefree nature of $\varepsilon_{\mu}^{\nu}$ gives

\begin{equation}
-\epsilon+\sigma_{r}+2\sigma_{t}=0
\end{equation}

Hence we have four algebraic equations (17)-(20) containing five
unknowns viz. $\rho$, $p$, $\epsilon$, $\sigma_{r}$ and $\sigma_{t}$.
In the following, we shall present exact solutions for various
assumptions:\\

\textbf{\underline{Case-I:}}~~~~~~~~~~~~~\underline{$\sigma_{t}=0$}\\

The complete solution is

\begin{equation}
\left.
\begin{array}{ll}
\rho=\lambda\left[\sqrt{1+\frac{2\mu}{\lambda r^{2}}}-1\right],~~\mu=\frac{5+3d}{32\pi}\\\\
p=\frac{\rho-\frac{1+d}{8\pi r^{2}}}{\sqrt{1+\frac{2\mu}{\lambda r^{2}}}}\\\\
\rho+p=\frac{3+d}{16\pi r^{2}\sqrt{1+\frac{2\mu}{\lambda r^{2}}}}>0\\\\
\epsilon=\sigma_{r}=\frac{5-d}{4r^{2}}
\end{array}
\right\}
\end{equation}

\textbf{\underline{Case-II:}}~~~~~~~~~~~~~\underline{$\sigma_{r}=0$}\\

The explicit solution reads

\begin{equation}
\left.
\begin{array}{ll}
\rho=\lambda\left[\sqrt{1-\frac{2\nu}{\lambda r^{2}}}-1\right]<0,~~\nu=\frac{5-3d}{16\pi}\\\\
p=\frac{\rho+\frac{3(1-d)}{16\pi r^{2}}}{\sqrt{1-\frac{2\nu}{\lambda r^{2}}}}\\\\
\rho+p=\frac{3d-7}{16\pi r^{2}\sqrt{1-\frac{2\nu}{\lambda r^{2}}}}<0\\\\
\epsilon=-\frac{(5-d)}{2r^{2}}\\\\
\sigma_{t}=-\frac{(5-d)}{4r^{2}}
\end{array}
\right\}
\end{equation}

\textbf{\underline{Case-III:}}~~~~~~~~~~~~~\underline{$\epsilon=0$}\\

Here the solution becomes

\begin{equation}
\left.
\begin{array}{ll}
\rho=\lambda\left[\sqrt{1+\frac{d}{4\pi\lambda r^{2}}}-1\right]\\\\
p=\frac{\rho-\frac{7d+1}{48\pi r^{2}}}{\sqrt{1+\frac{d}{4\pi\lambda r^{2}}}}\\\\
\rho+p=\frac{5d-1}{48\pi r^{2}\sqrt{1+\frac{d}{4\pi\lambda r^{2}}}}>0\\\\
\sigma_{r}=-2\sigma_{t}=\frac{5-d}{6r^{2}}
\end{array}
\right\}
\end{equation}

\textbf{\underline{Case-IV:}}~~~~~~~~~~~~~\underline{$p=0$ (\bf{dust})}\\

The solution is given by

\begin{equation}
\left.
\begin{array}{ll}
\rho=\frac{\lambda}{2}\left[1\pm\sqrt{1-\frac{1+3d}{4\pi\lambda r^{2}}}\right]\\\\
\epsilon=-\frac{1+7d}{4r^{2}}+6\pi\lambda\left[1\pm\sqrt{1-\frac{1+3d}{4\pi\lambda r^{2}}}\right]\\\\
\sigma_{r}=\frac{3(1-d)}{4r^{2}}+2\pi\lambda\left[1\pm\sqrt{1-\frac{1+3d}{4\pi\lambda r^{2}}}\right]\\\\
\sigma_{t}=-\frac{d+1}{2r^{2}}+2\pi\lambda\left[1\pm\sqrt{1-\frac{1+3d}{4\pi\lambda r^{2}}}\right]
\end{array}
\right\}
\end{equation}

\textbf{\underline{Case-V:}}~~~~~~~~~~~~~\underline{$\sigma_{r}=\sigma_{t}=\sigma$ (\bf{say})}\\

Then from equations (11) and (12), we must have\\

~~~~~~~~~~$d=5$ i.e, $\omega_{1}=1/5$ and $\omega_{2}=-1/5$\\

So the effective pressure on the brane becomes isotropic and we have
got three equations viz.

\begin{equation}
\rho\left(1+\frac{\rho}{2\lambda}\right)-\frac{\epsilon}{8\pi}=\frac{5}{8\pi r^{2}}
\end{equation}

\begin{equation}
p\left(1+\frac{\rho}{\lambda}\right)+\frac{\rho^{2}}{2\lambda}-\frac{\sigma}{8\pi}
=-\frac{1}{8\pi r^{2}}
\end{equation}

and

\begin{equation}
\epsilon=3\sigma
\end{equation}

with four unknowns $\rho$, $p$, $\epsilon$ and $\sigma$. Let us assume the
equation of state for brane matter as $p=\omega\rho$. Then we have the solution

\begin{equation}
\rho=\frac{\lambda}{2}\left(\frac{1-3\omega}{1+3\omega}\right)\left[1
\pm\sqrt{1-\frac{4(1+3\omega)}{\pi\lambda r^{2}(1-3\omega)^{2}}}\right],
~(\omega\neq\frac{1}{3})
\end{equation}

Further, for conformally flat bulk, $\varepsilon_{\mu\nu}\equiv0$ i.e,\\
$\epsilon=\sigma_{r}=\sigma_{t}=0$, and we have

\begin{equation}
\left.
\begin{array}{ll}
\rho=\lambda\left[\sqrt{1+\frac{5}{4\pi\lambda r^{2}}}-1\right]\\\\
p=\frac{\rho-\frac{3}{4\pi r^{2}}}{\sqrt{1+\frac{5}{4\pi\lambda r^{2}}}}\\\\
\text{with}~~~~\rho+p=\frac{1}{2\pi r^{2}\sqrt{1+\frac{5}{4\pi\lambda r^{2}}}}>0
\end{array}
\right\}
\end{equation}

In the present work, we have shown that a wormhole in Einstein gravity
can be considered also as a wormhole in brane scenario. Here we have
shown that the wormhole solution is possible for various matter distribution
on the brane. It is interesting to note that, although the effective matter
distribution violates the null energy condition, but the brane matter itself
does not violate the null energy condition [11] in general. Except for Case-II
(i.e, $\sigma_{r}=0$), in all other cases, the matter distribution obeys null
energy condition. However, Case-II is a peculiar one as energy density is
negative here. One may interprete such matter as ghost scalar field [17] or
tachyonic matter field on the brane. In cases IV and V, where matter in the
brane is of the form of dust or a perfect fluid with linear equation of state,
the solutions obtained are similar to those presented by Lobo [11] [see eqns.
(33)-(36) and (42)]. In solution (24), both the signs are possible as explicitly
discussed in [11]. the solution (28) is valid for $\omega<1/3$ (otherwise $\rho$
will be negative), but if $\omega$ is associated to the phantom era ($\omega<-1/3$),
then we have positive energy density only for the negative sign within the square
bracket. Therefore, instead of solving the complicated non linear second order
differential equation to obtain wormhole solution, we have assumed the solution
and obtained the possible matter distribution on the brane. For future work,
we shall extend this idea for various cosmological solutions on the brane.\\

{\bf Acknowledgement}:\\

The work has been done during a visit to IUCAA under the associateship programme.
The authors gratefully acknowledge the warm hospitality and facilities of work
at IUCAA. Also T.B is thankful to CSIR, Govt. of India, for awarding Junior Research
Fellowship.\\

\end{document}